\pdfoutput=1
\documentclass[twocolumn,pra,superscriptaddress,longbibliography]{revtex4-1}
\usepackage{graphicx,amsmath,amssymb}
\usepackage[colorlinks=true, citecolor=blue, urlcolor=blue, linkcolor=red]{hyperref}

\begin{document}
\title{Effects of counter-rotating-wave terms on the noisy frequency estimation}

\author{Ze-Zhou Zhang}
\affiliation{ Key Laboratory of Theoretical Physics of Gansu Province, \\
and Lanzhou Center for Theoretical Physics, Lanzhou University, Lanzhou, 730000, China}

\author{Wei Wu}
\email{wuw@lzu.edu.cn}
\affiliation{ Key Laboratory of Theoretical Physics of Gansu Province, \\
and Lanzhou Center for Theoretical Physics, Lanzhou University, Lanzhou, 730000, China}

\begin{abstract}
We investigate the problem of estimating the tunneling frequency of a two-level atomic system embedded in a dissipative environment by employing a numerically rigorous hierarchical equations of motion method. The effect of counter-rotating-wave terms on the attainable precision of the noisy quantum metrology is systematically studied beyond the usual framework of perturbative treatments. We find the counter-rotating-wave terms are able to boost the noisy quantum metrological performance in the intermediate and strong coupling regimes, whether the dissipative environment is composed of bosons or fermions. The result presented in this paper may pave a guideline to design a high-precision quantum estimation scenario under practical decoherence.
\end{abstract}
\maketitle

\section{Introduction}\label{sec:sec1}

Quantum parameter estimation is a rapidly developing research field, which has many potential applications from gravitational wave detection~\cite{PhysRevLett.123.231107,PhysRevLett.123.231108}, quantum radar~\cite{PhysRevLett.124.200503,PRXQuantum.2.030303}, quantum illumination~\cite{PhysRevLett.101.253601,PhysRevLett.127.040504} to various ultrasensitive quantum thermometries~\cite{PhysRevLett.114.220405,PRXQuantum.2.020322,PhysRevE.104.014136} and magnetometers~\cite{PhysRevX.10.011018,PhysRevA.99.062330}.  As demonstrated in many previous theoretical and experimental studies, using quantum coherence~\cite{2019}, quantum entanglement~\cite{PhysRevLett.79.3865,Nagata726,RevModPhys.90.035005}, quantum squeezing~\cite{PhysRevD.23.1693,PhysRevLett.123.040402,Chalopin2018} and quantum criticality~\cite{PhysRevA.78.042106,2014,PhysRevLett.121.020402,PhysRevLett.126.010502}, the performance of quantum metrology can surpass the so-called shot-noise limit or standard quantum limit, which is usually achieved in classical metrological schemes.

Unfortunately, the superiority of quantum metrology is commonly destroyed by decoherence~\cite{PhysRevLett.123.040402,PhysRevLett.109.233601,PhysRevA.94.042122,PhysRevApplied.5.014007,PhysRevLett.120.140501,2018Haase,Tamascelli_2020,PhysRevA.102.032607,PhysRevA.104.022612}, which is induced by the inevitable interaction with the surrounding environment. For example, Ref.~\cite{PhysRevLett.123.040402} reported that the Zeno-limit-type scaling relation, which is generated by the resource of quantum squeezing in the noiseless case, is degraded into the shot-noise limit under the influence of decoherence. In this sense, in any practical scheme, the decoherence should be carefully taken into account, which gives rise to the development of the so-called noisy quantum metrology~\cite{2018Haase}.

However, almost all the present studies of noisy quantum metrology restricted their attentions to the bosonic environment situation with the pure dephasing mechanism~\cite{PhysRevLett.109.233601,PhysRevLett.120.140501,PhysRevA.92.010302,PhysRevA.100.032318} or the rotating-wave approximation (RWA)~\cite{PhysRevA.88.035806,PhysRevA.96.052130,PhysRevApplied.15.054042}. On the other hand, the counter-rotating-wave terms play a significant role in the atomic spontaneous emission~\cite{PhysRevA.87.033818}, the quantum Zeno and anti-Zeno effect~\cite{PhysRevLett.101.200404,PhysRevA.81.042116} as well as the non-Markovianity in open quantum systems~\cite{PhysRevA.88.052111,PhysRevA.96.032125}. A question naturally raised here: what is the influence of the counter-rotating-wave terms on the estimation precision of a noisy quantum metrology? To address the above concern, with the help of the non-perturbative hierarchical equations of motion (HEOM) method~\cite{doi:10.1063/5.0011599,YAN2004216,PhysRevE.75.031107,doi:10.1063/1.2713104,doi:10.1063/1.2938087,PhysRevA.98.012110,PhysRevA.98.032116}, we go beyond the usual RWA treatment and investigate the performance of estimating the frequency of a two-level system which interacts with a bosonic or a fermionic environment.

This paper is organized as follows. In Sec.~\ref{sec:sec2}, we first recall some basic concepts, mainly about the classical and the quantum Fisher information, in the quantum parameter estimation theory. In Sec.~\ref{sec:sec3}, we propose our noisy quantum metrology scheme and outline our methodology. The effect of the counter-rotating-wave terms on our metrological precision is analyzed in Sec.~\ref{sec:sec4}. Some concerned discussions and the main conclusions of this paper are drawn in Sec.~\ref{sec:sec5}. In several appendixes, we provide some additional details about the HEOM method as well as the other two additional dynamical approaches. Throughout the paper, we set $\hbar=k_{\mathrm{B}}=1$.

\section{Quantum parameter estimation}\label{sec:sec2}

Generally, in a typical quantum metrology scheme, the parameter of interest, say $\theta$ in this section, is encoded into the state of a quantum system (acting as a probe) via a unitary or a nonunitary dynamical process. Then, the information about $\theta$ can be extracted from the $\theta$-dependent state $\varrho_{\theta}$ via repeated quantum measurements. In the above scenario, the metrological precision with respect to a given measurement scheme is constrained by the famous Cram$\mathrm{\acute{e}}$r-Rao bound~\cite{SMKay,PhysRevLett.72.3439,2018zhang}
\begin{equation}\label{eq:eq1}
\delta^{2}\theta\geq \frac{1}{\upsilon \mathcal{F}_{\text{C}}},
\end{equation}
where $\delta^{2}\theta$ denotes the variance of the derived estimator, $\upsilon$ is the number of repeated measurements (we set $\upsilon=1$ for the sake of consentience in this paper) and $\mathcal{F}_{\text{C}}$ is the classical Fisher information (CFI) corresponding to the selected measurement scheme. To perform such a measurement in the theory of quantum mechanism, one needs to construct a set of positive-operator valued measurement operators $\{\hat{\Pi}_{u}\}$, which satisfy $\sum_{u}\hat{\Pi}_{u}^{\dagger}\hat{\Pi}_{u}=\mathbf{\hat{1}}$ with discrete measurement outcomes $\{u\}$. For an arbitrary $\theta$-dependent state $\varrho_{\mathrm{\theta}}$, the measurement operator $\hat{\Pi}_{u}$ yields an outcome $u$ with a corresponding probability distribution $p(u|\theta)=\mathrm{Tr}(\hat{\Pi}_{u}\varrho_{\theta}\hat{\Pi}_{u}^{\dagger})$. With all the probabilities from $\{\hat{\Pi}_{u}\}$ at hand, the CFI can be computed via~\cite{SMKay,PhysRevLett.72.3439,2018zhang}
\begin{equation}\label{eq:eq2}
\mathcal{F}_{\text{C}}=\sum_{u}p(u|\theta)\bigg{[}\frac{\partial}{\partial\theta}\ln p(u|\theta)\bigg{]}^{2}.
\end{equation}

From Eq.~(\ref{eq:eq2}), one can find the value of CFI strongly relies on the choice of measurement operators. Running over all the possible measurement schemes, one can prove that the ultimate metrological precision is bounded by the following quantum Cram$\mathrm{\acute{e}}$r-Rao inequality~\cite{2019liu}
\begin{equation}\label{eq:eq3}
\delta^{2}\theta\geq \frac{1}{\upsilon \mathcal{F}_{\text{Q}}},
\end{equation}
where $\mathcal{F}_{\text{Q}}\equiv \mathrm{Tr}(\hat{\varsigma}^{2}\varrho_{\theta})$ with $\hat{\varsigma}$ determined by $\partial_{\theta}\varrho_{\theta}=\frac{1}{2}(\hat{\varsigma}\varrho_{\theta}+\varrho_{\theta}\hat{\varsigma})$ is the so-called quantum Fisher information (QFI). Specially, if $\varrho_{\theta}$ is a two-dimensional density matrix in the Bloch representation, namely, $\varrho_{\theta}=\frac{1}{2}(\mathbf{1}_{2}+\vec{r}\cdot\vec{\sigma})$ with $\vec{r}$ being the Bloch vector and $\vec{\sigma}\equiv(\hat{\sigma}_{x},\hat{\sigma}_{y},\hat{\sigma}_{z})$ being the vector of Pauli matrices, the QFI can be easily calculated via~\cite{2019liu,PhysRevA.87.022337}
\begin{equation}\label{eq:eq4}
\mathcal{F}_{\text{Q}}=|\partial_{\theta}\vec{r}|^{2}+\frac{(\vec{r}\cdot\partial_{\theta}\vec{r})^{2}}{1-|\vec{r}|^{2}}.
\end{equation}
If $\varrho_{\theta}$ is a pure state, Eq.~(\ref{eq:eq4}) reduces to $\mathcal{F}_{\text{Q}}=|\partial_{\theta}\vec{r}|^{2}$.

Physically speaking, the QFI describes the most statistical information in $\varrho_{\theta}$, while the CFI denotes the statistical message extracted from the selected measurement scheme. Thus, one can immediately conclude that $\mathcal{F}_{\text{Q}}\geq \mathcal{F}_{\text{C}}$. When the selected measurement scheme is the optimal one, $\mathcal{F}_{\text{C}}$ can be saturated to $\mathcal{F}_{\text{Q}}$. Unfortunately, there is no general way to find the optimal measurement operator, which has an explicit physical meaning. In this sense, designing the physically optimal measurement scheme, which can saturate the best attainable precision bounded by the QFI, is of importance in study of quantum metrology.

\section{Noisy quantum frequency estimation}\label{sec:sec3}

As discussed in several previous articles~\cite{PhysRevLett.109.233601,PhysRevLett.120.140501,PhysRevA.100.032318,PhysRevA.88.035806,Wang2017}, in the ideal noiseless case, the information of the estimated frequency, denoted by $\Delta$ in this paper, can be encoded to the state of a two-level atomic system, which acts as a quantum probe, via a unitary dynamics: $\varrho_{\Delta}=\varrho(t)=e^{-it\hat{H}_{\mathrm{s}}}\varrho(0)e^{it\hat{H}_{\mathrm{s}}}$ with $\hat{H}_{\text{s}}=\frac{1}{2}\Delta\hat{\sigma}_{x}$ being the Hamiltonian of the probe. However, such a unitary encoding scheme breaks down if the environmental influence is taken into account. In the noisy case, the encoding process is realized by a non-unitary dynamics $\varrho_{\Delta}=\varrho_{\text{s}}(t)=\text{Tr}_{\text{e}}[e^{-it\hat{H}}\varrho_{\text{se}}(0)e^{it\hat{H}}]$, where $\varrho_{\text{s}}(t)$ denotes the reduced density operator of the probe by tracing out the environmental degrees of freedom and $\hat{H}$ is the total Hamiltonian of the probe plus the environment.

\begin{figure*}
\centering
\includegraphics[angle=0,width=0.905\textwidth]{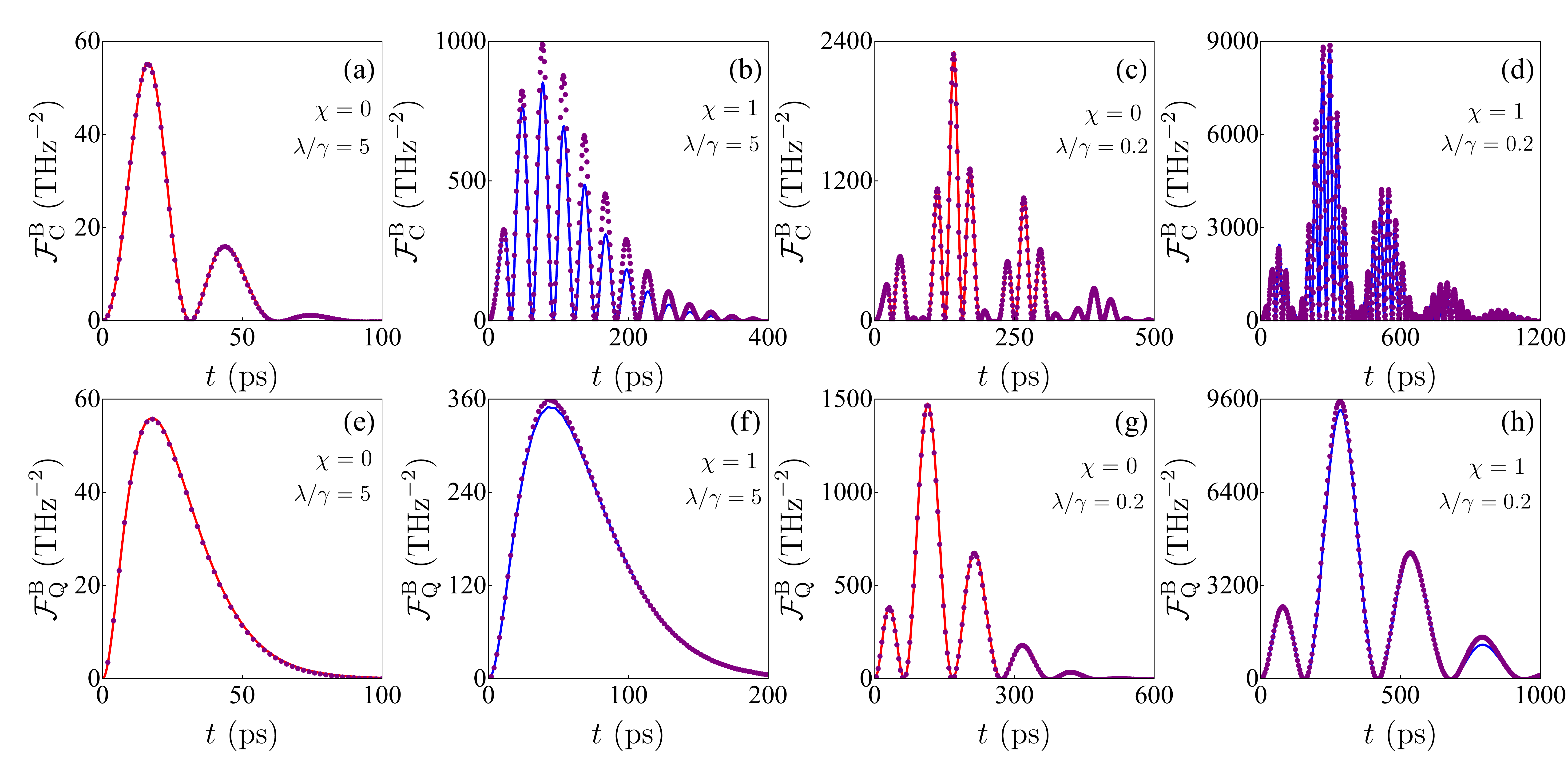}
\caption{The dynamics of the CFI (top panels) and the QFI (bottom panels) in Markovian regime ($\lambda/\gamma=5$) and non-Markovain regime ($\lambda/\gamma=0.2$). The red solid lines are analytical results from the RWA treatment, namely $\chi=0$. The blue solid lines are obtained by the Zwanzig-Nakajima master equation approach. The HEOM results are represented by purple circles. The coupling strength are chosen as (a) $\gamma=0.1~\mathrm{cm}^{-1}$, (b) $\gamma=0.02~\mathrm{cm}^{-1}$, (c) $\gamma=0.08~\mathrm{cm}^{-1}$, (d) $\gamma=0.04~\text{cm}^{-1}$ with $\Delta=0.1~\text{THz}$ and $\phi=\pi/4$, and (e) $\gamma=0.1~\mathrm{cm}^{-1}$, (f) $\gamma=0.04~\mathrm{cm}^{-1}$, (g) $\gamma=0.1~\mathrm{cm}^{-1}$, (h) $\gamma=0.04~\text{cm}^{-1}$ with $\Delta=10~\text{THz}$ and $\phi=\pi/4$.}\label{fig:fig1}
\end{figure*}

\begin{figure}
\centering
\includegraphics[angle=0,width=0.475\textwidth]{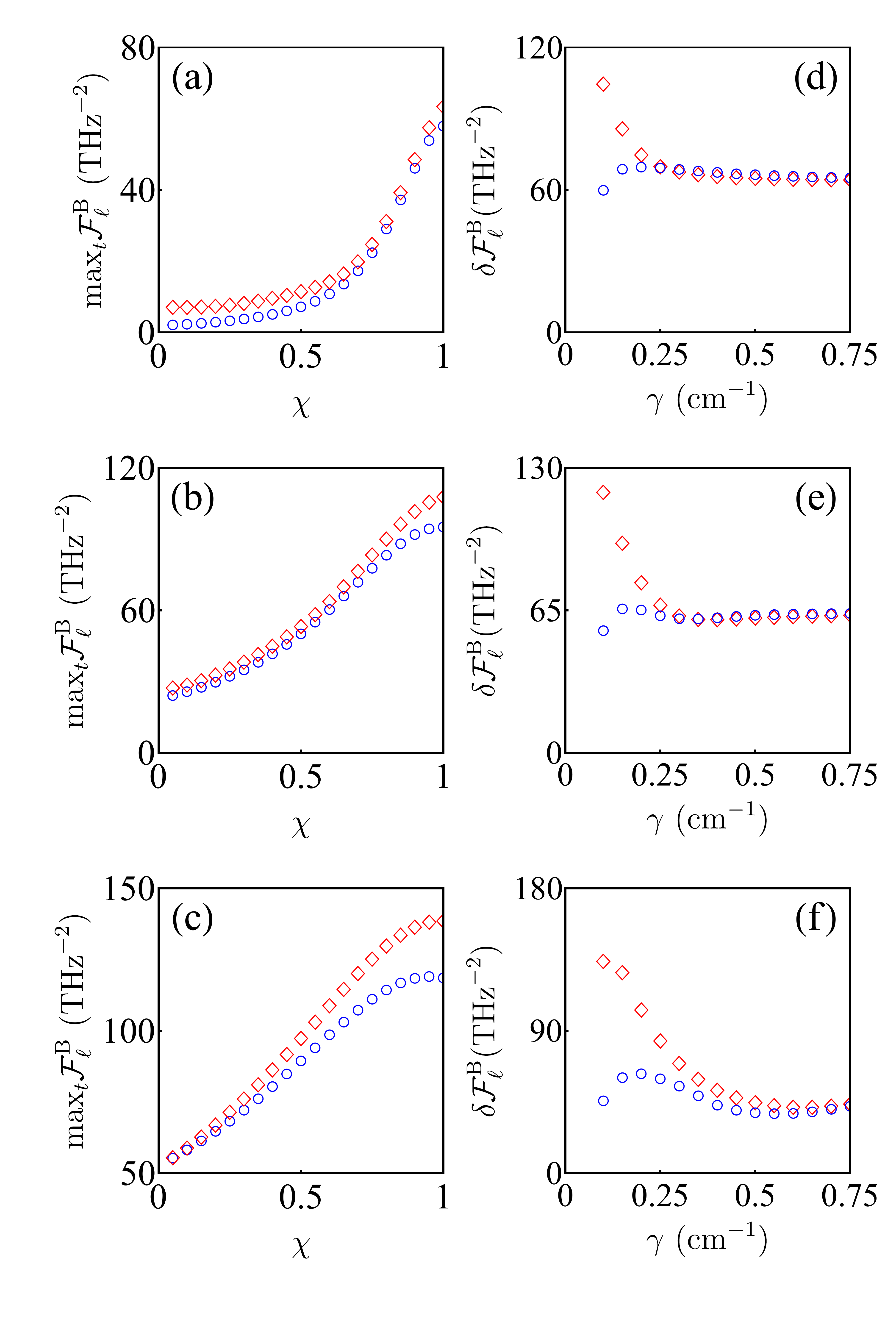}
\caption{Left panels: the optimal Fisher information with respect to the encoding time is plotted as a function of $\chi$ for the bosonic environment case. Right panels: the modification of Fisher information induced by the counter-rotating-wave terms is displayed as a function of $\gamma$ for the bosonic environment case. Parameters are chosen as (a) $\lambda/\gamma=0.8$, (b) $\lambda/\gamma=2$, (c) $\lambda/\gamma=3$ with $\lambda=0.32~\text{cm}^{-1}$, (d) $\lambda/\gamma=1.5$, (e) $\lambda/\gamma=1$, (f) $\lambda/\gamma=0.5$ with $\chi=1$. The red diamonds and the blue circles are, respectively, the QFI ($\ell=\text{Q}$) and the CFI ($\ell=\text{C}$) predicted by the HEOM method. Here, we mainly focus on non-Markovin and strong-coupling regimes in which the effect of counter-rotating-wave terms is significant. The other parameters are $\Delta=0.1~\text{THz}$ and $\phi=\pi/4$.}\label{fig:fig2}
\end{figure}

In this paper, we assume $\hat{H}$ has the following general form~\cite{PhysRevA.96.032125,PhysRevB.100.115106,Zhang:19,PhysRevB.101.155147}
\begin{equation}\label{eq:eq5}
\begin{split}
\hat{H}=&\hat{H}_{\text{s}}+\sum_{k}\omega_{k}\hat{c}_{k}^{\dag}\hat{c}_{k}+\sum_{k}\Big{(}g_{k}^{*}\hat{\sigma}_{-}\hat{c}_{k}^{\dag}+g_{k}\hat{\sigma}_{+}\hat{c}_{k}\Big{)}\\
&+\chi\sum_{k}\Big{(}g_{k}^{*}\hat{\sigma}_{+}\hat{c}_{k}^{\dag}+g_{k}\hat{\sigma}_{-}\hat{c}_{k}\Big{)},
\end{split}
\end{equation}
where $\hat{c}_{k}$ and $\hat{c}_{k}^{\dagger}$ are the annihilation and creation operators of the $k$th environmental mode with frequency $\omega_{k}$, respectively. If $\hat{c}_{k}$ and $\hat{c}_{k}^{\dagger}$ satisfy the canonical commutation relations $[\hat{c}_{k},\hat{c}_{k'}^{\dagger}]=\delta_{kk'}$, one can deem that the environment is composed of bosons. On the other hand, if $\hat{c}_{k}$ and $\hat{c}_{k}^{\dagger}$ obey $\{\hat{c}_{k},\hat{c}_{k'}^{\dagger}\}=\delta_{kk'}$, then the environment is a fermionic environment. Operators $\hat{\sigma}_{\pm}$ are defined as $\hat{\sigma}_{-}=\hat{\sigma}_{+}^{\dag}\equiv|-\rangle\langle +|$ with $|\pm\rangle$ being the eigenstates of Pauli-$x$ operator, i.e. $\hat{\sigma}_{x}|\pm\rangle=\pm|\pm\rangle$, and $g_{k}$ are complex numbers quantifying the coupling strength between the quantum probe and the $k$th environmental mode. The dimensionless parameter $\chi$ is a real number, satisfying $\chi\in[0,1]$, characterizes the weight of counter-rotating-wave terms in the interaction Hamiltonian. When $\chi=0$, all the counter-rotating-wave terms are removed and $\hat{H}$ reduces to the damped Jaynes-Cummings model~\cite{PhysRevA.88.035806,PhysRevA.96.052130,PhysRevApplied.15.054042,Wang2017}. When $\chi=1$, all the contributions of the counter-rotating-wave terms are taken into consideration and this Hamiltonian is totally beyond the RWA. By introducing such a parameter, a bridge between the RWA regime and the non-RWA regime can be built, which helps us to get a deeper understanding of the effect of counter-rotating-wave terms on the parameter estimation in a dissipative environment.

To obtain the information of the estimated frequency, one need to monitor the reduced density operator of the probe, namely $\varrho_{\text{s}}(t)$. In this paper, we adopt the HEOM method~\cite{doi:10.1063/5.0011599,YAN2004216,PhysRevE.75.031107,doi:10.1063/1.2713104,doi:10.1063/1.2938087,PhysRevA.98.012110,PhysRevA.98.032116}, which can provide a numerically rigours result, to solve this problem (see Appendix A for more details). As benchmarks, we also employ the other two approaches, the Zwanzig-Nakajima master equation technique (see Refs.~\cite{PhysRevB.71.035318,PhysRevB.79.125317} and Appendix B) and the RWA treatment (see Appendix C), to investigate the reduced dynamics of the probe. Together with the purely numerical results from the HEOM, these two additional methods can make our conclusion more persuasive.

In this paper, we assume the initial state of the whole system is given by
\begin{equation}\label{eq:eq6}
\varrho_{\text{se}}(0)=\varrho_{\text{s}}(0)\otimes\varrho_{\text{e}}=|\psi_{\text{s}}(0)\rangle\langle\psi_{\text{s}}(0)|\bigotimes_{k}|0_{k}\rangle\langle 0_{k}|,
\end{equation}
where $|\psi_{\text{s}}(0)\rangle=\cos\phi|+\rangle+\sin\phi|-\rangle$, and $|0_{k}\rangle$ denotes the vacuum state of the $k$th environmental mode. The spectral density of the environment $J(\omega)$, which is defined by $J(\omega)=\sum_{k}|g_{k}|^{2}\delta(\omega-\omega_{k})$, has a Lorentzian form
\begin{equation}\label{eq:eq7}
J(\omega)=\frac{1}{2\pi}\frac{\gamma\lambda^{2}}{(\omega-\Delta)^{2}+\lambda^{2}},
\end{equation}
where $\lambda$ defines the spectral width, $\gamma$ can be approximately interpreted as the probe-environemt coupling strength. The Lorentzian spectral density has a clear Markovian-non-Markovian boundary~\cite{Breuer,PhysRevLett.99.160502,PhysRevA.95.042132,PhysRevA.104.022612} and can give rise to an Ornstein-Uhlenbeck-type environmental correlation function, which guarantees the feasibility of the HEOM algorithm (see Appendix A for details).

As long as $\varrho_{\text{s}}(t)$ is obtained, the CFI and the QFI can be calculated by making use of Eq.~(\ref{eq:eq2}) and Eq.~(\ref{eq:eq4}), respectively. During the purely numerical calculations, one needs to handle the first-order derivative to the parameter $\Delta$, say $\partial_{\Delta}\vec{r}$ in Eq.~(\ref{eq:eq4}). In this paper, the first-order derivative for an arbitrary $\Delta$-dependent function $f_{\Delta}$ is numerically done by adopting the following finite difference method
\begin{equation}\label{eq:eq8}
\partial_{\Delta} f_{\Delta}\simeq\frac{-f_{\Delta+2\delta}+8f_{\Delta+\delta}-8f_{\Delta-\delta}+f_{\Delta-2\delta}}{12\delta}.
\end{equation}
We set $\delta/\Delta=10^{-6}$, which provides a very good accuracy.

\section{Results}\label{sec:sec4}

\begin{figure}
\centering
\includegraphics[angle=0,width=0.475\textwidth]{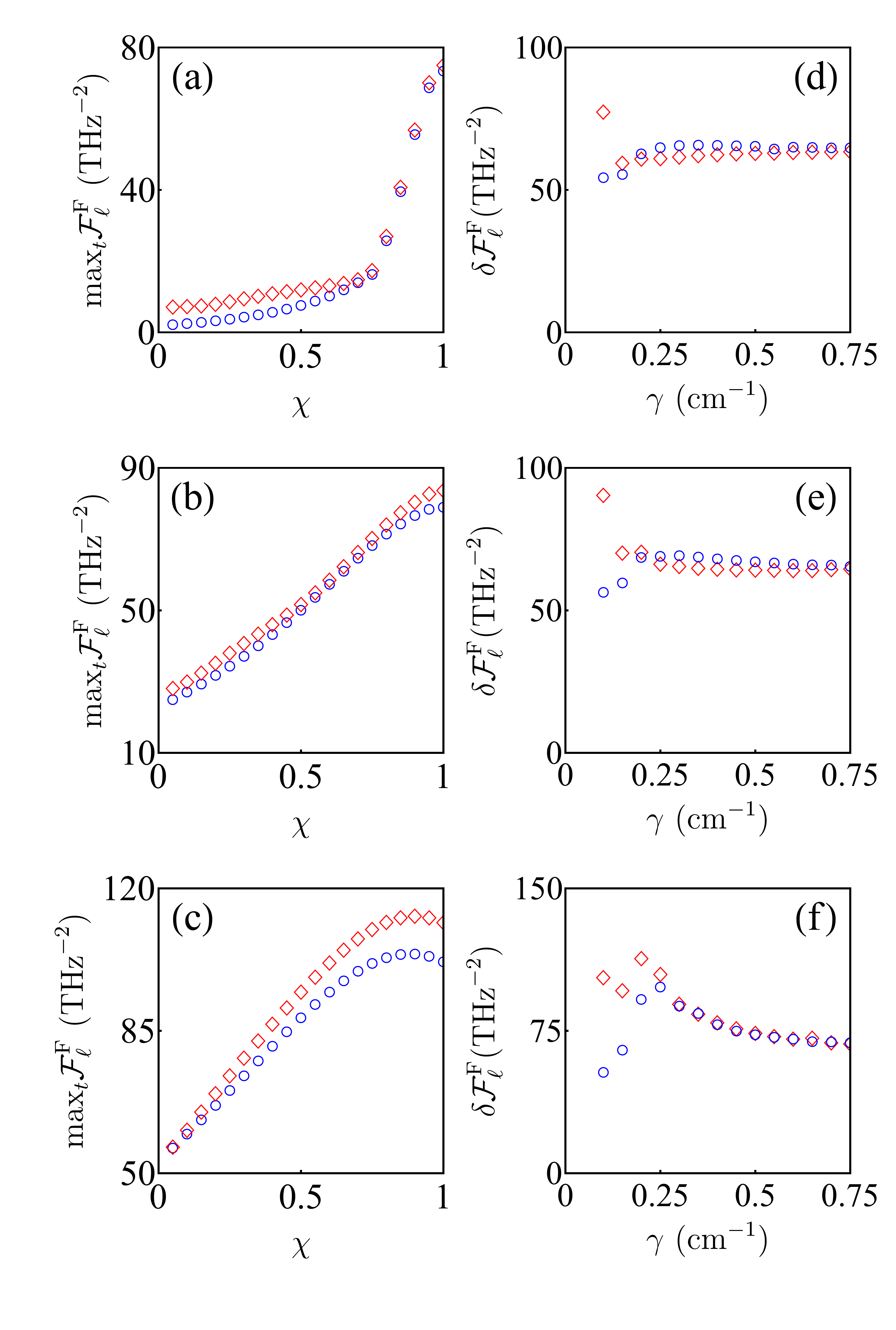}
\caption{The same with Fig.~\ref{fig:fig2}, but in the fermionic environment case.}\label{fig:fig3}
\end{figure}

In this section, we try to address the effect of counter-rotating-wave terms on the noisy quantum frequency estimation. To obtain the CFI, the measurement operators in this paper are chosen as $\hat{\Pi}=\{|e\rangle\langle e|,|g\rangle\langle g|\}$ with $|e,g\rangle$ being the eigenstates of Pauli-$z$ operator. In the ideal noiseless case, the CFI with respect to the selected measurement scheme is $\mathcal{F}_{\text{C}}^{\text{ideal}}=t^{2}$, which is completely saturated to the QFI (see Appendix C). This result means the choice of the measurement scheme is the optimal one. Moreover, our measurement scheme is mathematically equivalent to perform a standard Ramsey spectroscopy~\cite{PhysRevLett.79.3865,PhysRevLett.109.233601,Wang2017}, which is commonly employed to estimate the phases (or an external magnetic field) of cold atomic systems. As the detailed expositions of a typical Ramsey interferometer setup for a two-level system~\cite{PhysRevLett.79.3865,PhysRevLett.109.233601,Wang2017}, the uncertainty of estimating the parameter $\Delta$ from the error propagation formula is in full accord with that of the CFI obtained from Eq.~(\ref{eq:eq2}). Thus, it is quite natural to generalize the above measurement scheme to the noisy case and evaluate the environmental influences on its performance. We find, in the noisy situation, although the selected measurement scheme cannot completely saturate the ultimate bound given by the quantum Cram$\mathrm{\acute{e}}$r-Rao theorem (see the numerical results in Fig.~\ref{fig:fig2} (a-c) and Fig.~\ref{fig:fig3} (a-c)), the corresponding CFI is in qualitative agreement with the QFI in the non-Markovin and strong-coupling regimes.

\subsection{The bosonic environment case}\label{sec:sec4a}

In Fig.~\ref{fig:fig1}, we plot the dynamics of the CFI and the QFI for $\chi=0$ and $\chi=1$ including both the Markovian ($\lambda/\gamma\rightarrow \infty$) and the non-Markovian ($\lambda/\gamma\rightarrow 0$) cases. From Fig.~\ref{fig:fig1}, one can see both the CFI and the QFI gradually increase from zero to their maximum values with certain oscillations. Such oscillations are quite dramatic in non-Markovian regimes and may be linked to the information's backflow from the environment back to the probe~\cite{PhysRevA.95.042132,PhysRevA.88.035806,PhysRevA.104.022612}. As the encoding time becomes longer, both the CFI and the QFI begin to decrease and eventually vanish in the long-encoding-time limit. These results can be physically understood as the competition between the encoding process and the decoherence in the noisy quantum metrology~\cite{PhysRevA.102.032607,PhysRevA.104.022612,PhysRevApplied.15.054042}. At the beginning, the non-unitary encoding process generates the information of $\Delta$ in $\varrho_{\mathrm{s}}(t)$ which leads to the increase of the CFI (QFI) from zero. However, as the encoding time becomes longer, the message about $\Delta$ inevitably leaks to the environment and eventually destroyed by the decoherence. Our result suggests there exists an optimal encoding time which can maximize the value of CFI (QFI). In Fig.~\ref{fig:fig2} (a-c), we plot the maximal CFI (QFI) with respect to the encoding time as a function of $\chi$. One can see the optimal CFI (QFI) monotonously increases as the value of $\chi$ becomes large, which means the counter-rotating-wave terms can effectively boost the performance our noisy quantum metrology in the strong-coupling regime.

\begin{table*}[htp]
\begin{center}
\caption{The Fisher information (QFI and CFI) versus the coupling strength in the weak-coupling regime. Here, $\mathbb{R}_{\text{B}}$ is defined as $\mathbb{R}_{\ell}^{\text{B}}\equiv\max_{t}\mathcal{F}_{\ell}^{\mathrm{RWA}}/\max_{t}\mathcal{F}_{\ell}^{\mathrm{B}}$. Other parameters are the same with these of Fig.~\ref{fig:fig2}.}\label{table:table1}
\setlength{\tabcolsep}{7.75pt}
\begin{tabular}{ccccccccccc}
\hline
  \hline
  $\gamma~(\text{cm}^{-1})$ & 0.00 & 0.01 & 0.02 & 0.03 & 0.04 & 0.05 & 0.06 & 0.07 & 0.08 & 0.09 \\
  $\mathbb{R}_{\text{Q}}^{\text{B}}~(\lambda/\gamma=1.5)$ & 1.0000 & 0.9733 & 0.9294 & 0.9147 & 0.9524 & 0.8591 & 0.7122& 0.6115 & 0.5381& 0.4818 \\
  $\mathbb{R}_{\text{Q}}^{\text{B}}~(\lambda/\gamma=1.0)$ & 1.0000 & 0.9536 & 0.9286 & 0.9171 & 0.9009 & 0.9467 & 0.7775& 0.6622 & 0.5792& 0.5169 \\
  $\mathbb{R}_{\text{Q}}^{\text{B}}~(\lambda/\gamma=0.5)$ & 1.0000 & 0.9653 & 0.9286 & 0.9224 & 0.8786 & 0.8846 & 0.9050 & 0.8349 & 0.7205 & 0.6344 \\
  $\mathbb{R}_{\text{C}}^{\text{B}}~(\lambda/\gamma=1.5)$ & 1.0000 & 1.0164 & 1.0669 & 1.1262 & 1.2134 & 1.0619 & 0.8243 & 0.7631 & 0.6994 & 0.6355 \\
  $\mathbb{R}_{\text{C}}^{\text{B}}~(\lambda/\gamma=1.0)$ & 1.0000 & 1.0103 & 1.0390 & 1.0719 & 1.1493 & 1.2190 & 0.8633 & 0.8135 & 0.7597 & 0.7038 \\
  $\mathbb{R}_{\text{C}}^{\text{B}}~(\lambda/\gamma=0.5)$ & 1.0000 & 1.0045 & 1.0176 & 1.0410 & 1.0535 & 1.1140 & 1.1744 & 1.0459 & 0.8518 & 0.8137 \\
  \hline
  \hline
\end{tabular}
\end{center}
\end{table*}

On the other hand, we define the following quantities
\begin{equation}\label{eq:eq9}
\delta \mathcal{F}_{\ell}^{\text{B}}\equiv \max_{t}\mathcal{F}_{\ell}^{\text{B}}-\max_{t}\mathcal{F}_{\ell}^{\mathrm{RWA}},
\end{equation}
to quantify the effect of counter-rotating-wave terms on the metrological precision. Here, the superscript $\ell=\text{C}$ or $\text{Q}$ implies the CFI or the QFI, $\mathcal{F}_{\ell}^{\text{B}}$ and $\mathcal{F}_{\ell}^{\mathrm{RWA}}$ denote the Fisher information for $\chi=1$ and $\chi=0$, respectively. As long as $\delta \mathcal{F}_{\ell}^{\text{B}}>0$, one can conclude that the counter-rotating-wave terms play a positive role in improving the metrological precision. In Fig.~\ref{fig:fig2} (d-f), we plot $\delta \mathcal{F}_{\ell}^{\text{B}}$ as a function of the coupling constant $\gamma$. As one can see from the Fig.~\ref{fig:fig2} (d-f), $\delta \mathcal{F}_{\ell}^{\text{B}}$ remains positive in the entire range of intermediate and strong coupling regimes (say, $0.10~\text{cm}^{-1}\leq\gamma\leq 0.75~\text{cm}^{-1}$ in Fig.~\ref{fig:fig2}). In Table~\ref{table:table1} and Fig.~\ref{fig:figadd}(c), we show the Fisher information with and without RWA in the weak-coupling regime, which demonstrates $\max_{t}\mathcal{F}_{\text{Q}}^{\text{B}}>\max_{t}\mathcal{F}_{\text{Q}}^{\mathrm{RWA}}$ as well. These results mean the consideration of counter-rotating-wave terms can provide a larger QFI. However, we also notice that $\max_{t}\mathcal{F}_{\text{C}}^{\text{B}}$ can be smaller than $\max_{t}\mathcal{F}_{\text{C}}^{\mathrm{RWA}}$ when $\gamma$ is very small. Such a deviation from that of the QFI situation is probably induced by the fact that the selected measurement scheme $\hat{\Pi}=\{|e\rangle\langle e|,|g\rangle\langle g|\}$ is not the optimal one in the noisy environment. This implies the CFI can not be completely aligned with the behaviour of QFI.

\subsection{The fermionic environment case}\label{sec:sec4b}

\begin{table*}[htp]
\begin{center}
\caption{The same with Table.~\ref{table:table1}, but in the fermionic environment case. Here, $\mathbb{R}_{\text{F}}$ is defined as $\mathbb{R}_{\ell}^{\text{F}}\equiv\max_{t}\mathcal{F}_{\ell}^{\mathrm{RWA}}/\max_{t}\mathcal{F}_{\ell}^{\mathrm{F}}$.}\label{table:table2}
\setlength{\tabcolsep}{7.75pt}
\begin{tabular}{cccccccccccc}
\hline
  \hline
  $\gamma~(\text{cm}^{-1})$ & 0.00 & 0.01 & 0.02 & 0.03 & 0.04 & 0.05 & 0.06 & 0.07 & 0.08 & 0.09 \\
  $\mathbb{R}_{\text{Q}}^{\text{F}}~(\lambda/\gamma=0.5)$ & 1.0000 & 0.9657 & 0.9341 & 0.9326 & 0.8846 & 0.9079 & 1.0435 & 0.8907 & 0.7991 & 0.6955 \\
  $\mathbb{R}_{\text{Q}}^{\text{F}}~(\lambda/\gamma=1.0)$ & 1.0000 & 0.9577 & 0.9411 & 0.9383 & 0.9760 & 1.0099 & 0.8440 & 0.7301 & 0.6476 & 0.5854 \\
  $\mathbb{R}_{\text{Q}}^{\text{F}}~(\lambda/\gamma=1.5)$ & 1.0000 & 0.9733 & 0.9655 & 0.9720 & 1.1567 & 0.9321 & 0.7861 & 0.6842 & 0.6119 & 0.5558 \\
  $\mathbb{R}_{\text{C}}^{\text{F}}~(\lambda/\gamma=0.5)$ & 1.0000 & 1.0019 & 1.0058 & 1.0151 & 1.0036 & 1.0286 & 1.0846 & 1.0761 & 0.8291 & 0.7891 \\
  $\mathbb{R}_{\text{C}}^{\text{F}}~(\lambda/\gamma=1.0)$ & 1.0000 & 1.0094 & 1.0298 & 1.0438 & 1.1307 & 1.2738 & 0.8496 & 0.8008 & 0.7496 & 0.6975 \\
  $\mathbb{R}_{\text{C}}^{\text{F}}~(\lambda/\gamma=1.5)$ & 1.0000 & 1.0234 & 1.0785 & 1.1280 & 1.3398 & 1.0534 & 0.8209 & 0.7654 & 0.7085 & 0.6520 \\
  \hline
  \hline
\end{tabular}
\end{center}
\end{table*}

\begin{figure}
\centering
\includegraphics[angle=0,width=0.480\textwidth]{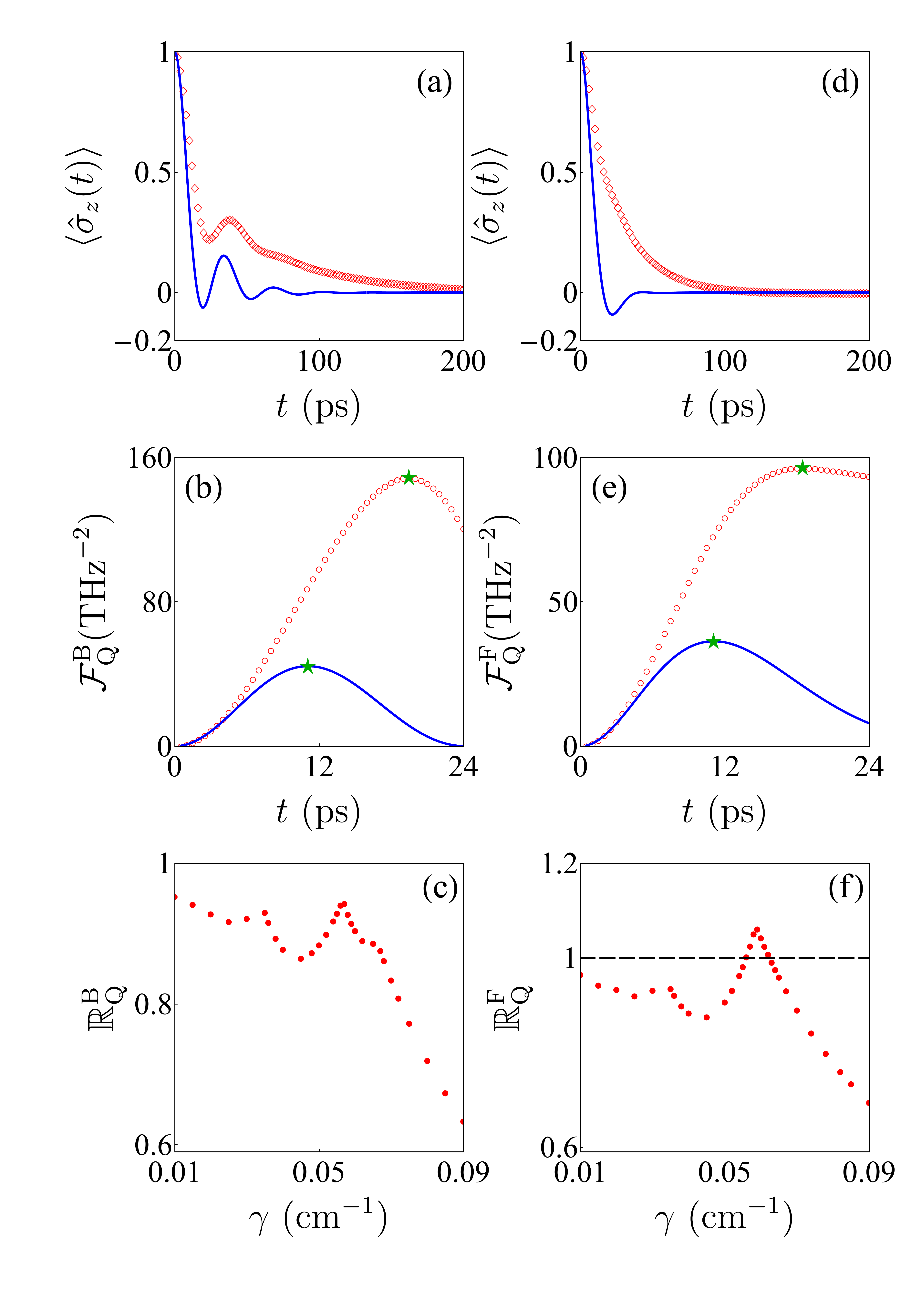}
\caption{The dynamics of the population difference $\langle\hat{\sigma}_{z}(t)\rangle$ with (blue solid lines) and without (red rhombus) RWA in the strong-coupling regime for (a) the bosonic environment case ($\gamma=0.2~\text{cm}^{-1}$ and $\lambda/\gamma=0.5$) and (d) the fermionic environment case ($\gamma=0.15~\text{cm}^{-1}$ and $\lambda/\gamma=1.5$). The dynamics of the QFI $\mathcal{F}_{\text{Q}}(t)$ with (blue solid lines) and without (red circles) RWA in the strong-coupling regime for (b) the bosonic environment case ($\gamma=0.2~\text{cm}^{-1}$ and $\lambda/\gamma=0.5$) and (e) the fermionic environment case ($\gamma=0.15~\text{cm}^{-1}$ and $\lambda/\gamma=1.5$). Here, the green five-point stars mark the positions of the optimal QFIs with respect the encoding time. The ratio of $\mathbb{R}_{\ell}$ versus $\gamma$ in the weak-coupling regime with $\lambda/\gamma=0.5$ for (c) the bosonic environment case and (f) the fermionic environment case. The other parameters are chosen as $\Delta=0.1~\text{THz}$ and $\phi=\pi/4$.}\label{fig:figadd}
\end{figure}

Almost all the existing studies of the noisy quantum metrology have restricted their attention to the bosonic environment case. In this subsection, we generalize our study to the fermionic environment case with the help of the HEOM method (see Appendix A). We find the Fisher information in the fermionic environment has a similar dynamical behavior with that of the bosonic environment case (not shown here). Thus, by optimizing the encoding time, the maximal CFI and QFI versus $\chi$ can be obtained. As displayed in Fig.~\ref{fig:fig3} (a-c), one can see the metrological precision can be improved by increasing the weight of the counter-rotating-wave terms. Similar to the bosonic environment case, we here define
\begin{equation}\label{eq:eq10}
\delta \mathcal{F}_{\ell}^{\text{F}}\equiv \max_{t}\mathcal{F}_{\ell}^{\text{F}}-\max_{t}\mathcal{F}_{\ell}^{\mathrm{RWA}},
\end{equation}
as the quantity to detect the influence of the counter-rotating-wave terms on our metrological precision in the fermionic environment case with a superscript $\text{F}$. As displayed in Fig.~\ref{fig:fig3} (d-f) and Table ~\ref{table:table2}, we confirm that $\delta \mathcal{F}_{\ell}^{\text{F}}$ is positive in the strong-coupling regime for the fermionic environment case. These results demonstrate that the counter-rotating-wave terms are able to boost the QFI by enhancing the probe-environment coupling, whether the environment is composed of bosons or fermions. However, if $\gamma$ is very small, an anomalous phenomenon of $\mathbb{R}_{\ell}^{\text{F}}>1$ may occurs (see Table ~\ref{table:table2} as well as Fig.~\ref{fig:figadd}(f)).

Here, we would like to provide a possible physical explain on why the counter-rotating-wave terms fail to improve the metrological performance in small-$\gamma$ regime. Essentially speaking, we concentrate on the problem of noisy quantum parameter estimation, which means the decoherence induced by the system-environment coupling plays a negative role in the metrological performance. It is commonly thought that a more severe decoherence always induces a lower sensitivity, because a stronger noise leads to a more serious damage to the ideal precision obtained in noiseless case. On the other hand, the system-environment interaction with and without RWA can be viewed as two completely different decoherence channels. Such a difference gives rise to distinct decay rates as well as disparate estimation accuracies. As displayed in Fig.~\ref{fig:figadd}(a) and Fig.~\ref{fig:figadd}(d), we find, in the strong-coupling regime, the inclusion of counter-rotating-wave terms can effectively inhibit decoherence. This decoherence-suppression effect was also reported in Ref.~\cite{PhysRevA.90.022122} and can lead to a larger $\max_{t}\mathcal{F}_{\ell}$ (see Fig.~\ref{fig:figadd}(b) and Fig.~\ref{fig:figadd}(e)). However, as the coupling strength decreases, the influence of counter-rotating-wave terms vanishes, resulting in the above mechanism of decoherence-suppression breaks down. Thus, in the weak-coupling regime, the phenomenon of counter-rotating-wave-terms-boosted quantum metrology may disappear. Such an anomalous phenomenon is displayed in Fig.~\ref{fig:figadd}(f), from which one can see the ratio of $\mathbb{R}_{\text{Q}}^{\text{F}}$ can be larger than one if $\gamma$ is small.

\section{Discussion and Conclusion}\label{sec:sec5}

Before concluding our work, some important remarks shall be addressed here.

(1) In our work, the strength of the counter-rotating-wave terms in the interaction Hamiltonian, namely the parameter $\chi$, is continuously tunable for the entire range of $\chi\in[0,1]$, which can build a bridge connecting two particular limits: totally with and without RWA. In fact, such a controllable strength of the counter-rotating-wave terms in the quantum Rabi model is called anisotropy~\cite{PhysRevA.95.043823,PhysRevLett.112.173601,PhysRevLett.119.220601}, which leads to a much richer ground-state phase diagram~\cite{PhysRevLett.112.173601,PhysRevLett.119.220601}, and can be used to explain the anomalous Bloch-Siegert shift in the ultrastrong-coupling regime~\cite{PhysRevX.4.021046}. These previous studies of the anisotropic quantum Rabi model motivate us to explore the influence of the counter-rotating-wave terms on the metrological precision by introducing the engineered tunable parameter $\chi$.

(2) Compared with the widely acceptable interaction Hamiltonian between a two-level spin and a bosonic environment, the form of spin-fermion interaction still remains controversial. The spin-fermion model considered in this paper is the most straightforward generalization of the well-accepted spin-boson model and can be regarded as a toy model. However, it is necessary to emphasize that such a spin-fermion model is not merely of academic interest. It may be used to describe certain physical phenomena, say, the magnetism of the weak-moment heavy-fermion compound $\text{UR}_{2}\text{Si}_2$~\cite{PhysRevB.54.9322,doi:10.1063/1.2712944}.

(3) Though the theoretical model and the numerical tool are the same with these of Ref.~\cite{PhysRevA.96.032125}, the centered goal of our work is utterly different from that of Ref.~\cite{PhysRevA.96.032125}. Our paper concentrate on a concrete physical problem, while the Ref.~\cite{PhysRevA.96.032125} is much closer to a methodology article discussing the HEOM method. What's more important, we here investigate an alternative topic, which means the physical conclusion in Ref.~\cite{PhysRevA.96.032125} cannot be straightforwardly applied to our present paper.

(4) Simulating an exact non-Markovian dissipative dynamics of an open quantum system in the strong-coupling regime is difficult. As reported in Refs.~\cite{PhysRevLett.109.170402,doi:10.1021/acsphotonics.8b01455,PhysRevE.90.022122,PhysRevA.93.020105,Ran_on_2013,PhysRevLett.115.168901}, the so-called negative frequency phenomenon may appear in the strong-coupling regime, which leads to isolated eigenenergies in the energy spectrum (the bound state effect)~\cite{doi:10.1021/acsphotonics.8b01455,PhysRevE.90.022122,PhysRevA.93.020105}. The appearance of negative frequencies can lead to a long-lived coherence~\cite{PhysRevLett.109.170402,doi:10.1021/acsphotonics.8b01455,PhysRevA.93.020105} as well as the breakdown of canonical thermalization~\cite{PhysRevE.90.022122,PhysRevLett.115.168901}, which cannot be predicted by the common quantum master equation approach. However, to the best of our knowledge, such a negative frequency phenomenon usually occurs in the unbiased spin-boson model with RWA~\cite{doi:10.1021/acsphotonics.8b01455,PhysRevA.93.020105} or the quantum Brownian motion (the Caldeira-Leggett model) without adding the counter term~\cite{Ran_on_2013,PhysRevLett.115.168901}. For the model considered in this paper (Lorentzian spin-boson model totally beyond the RWA), the negative frequency phenomenon has not been reported. Excluding the negative frequency problem, we believe the numerical correctness from the HEOM method can be fully guaranteed.

In summary, using the numerically rigorous HEOM method, we investigate the exact reduced dynamics of two-level system coupled to a dissipative bosonic or fermionic environment. With these results, we analyze the influence of counter-rotating-wave terms on the performance of a noisy parameter estimation beyond the usual weak-coupling treatment. It is revealed that the counter-rotating-wave terms can effectively enhance the metrological precision in the non-Markovian and strong-coupling regimes, whether the dissipative environment is composed of bosons or fermions. Our results may have certain applications in quantum metrology and quantum sensing.

\section*{Acknowledgments}

The authors wish to thank Dr. S.-Y. Bai, Dr. C. Chen, Prof. H.-G. Luo and Prof. J.-H. An for many useful discussions. The work was supported by the National Natural Science Foundation (Grant No. 11704025 and No. 12047501).

\section*{Appendix A: the HEOM method}

In this appendix, we briefly sketch the HEOM method following the detailed expositions in Refs.~\cite{PhysRevA.96.032125,PhysRevA.98.012110,PhysRevA.98.032116,Suess2015}. Let us consider a general quantum dissipative system whose Hamiltonian can be described by
\setcounter{equation}{0}
\renewcommand\theequation{A\arabic{equation}}
\begin{equation}\label{eq:eqa1}
\hat{H}=\hat{H}_{\text{s}}+\sum_{k}\omega_{k}\hat{c}_{k}^{\dag}\hat{c}_{k}+\sum_{k}\Big{(}g_{k}^{*}\hat{L}\hat{c}_{k}^{\dag}+g_{k}\hat{L}^{\dagger}\hat{c}_{k}\Big{)},
\end{equation}
where $\hat{L}$ is the quantum probe's operator coupled to its surrounding dissipative environment. Taking $\hat{L}=\hat{\sigma}_{-}+\chi\hat{\sigma}_{+}$, Eq.~(\ref{eq:eq5}) in the main text can be recovered. The dynamics of the Hamiltonian $\hat{H}$ is governed by the common Schr$\ddot{\mathrm{o}}$dinger equation. By introducing the bosonic or the fermionic coherent-state $|\textbf{z}\rangle\equiv\bigotimes_{k}|z_{k}\rangle$ with $\hat{c}_{k}|z_{k}\rangle=z_{k}|z_{k}\rangle$, one can recast the standard Schr$\ddot{\mathrm{o}}$dinger equation into the following non-Markovian quantum state diffusion~\cite{PhysRevLett.105.240403,PhysRevLett.113.200403,PhysRevLett.113.150403}
\begin{equation}\label{eq:eqa2}
\begin{split}
\frac{\partial}{\partial t}|\psi_{t}(\textbf{z}^{*})\rangle=&-i\hat{H}_{\text{s}}|\psi_{t}(\textbf{z}^{*})\rangle+\hat{L}\textbf{z}_{t}^{*}|\psi_{t}(\textbf{z}^{*})\rangle\\
&-\hat{L}^{\dagger}\int_{0}^{t}d\tau C(t-\tau)\frac{\delta}{\delta \textbf{z}_{\tau}^{*}}|\psi_{t}(\textbf{z}^{*})\rangle,
\end{split}
\end{equation}
where $|\psi_{t}(\textbf{z}^{*})\rangle$ is the wave function in the coherent-state representation. The stochastic variable (random noise) $\textbf{z}_{t}\equiv i\sum_{k}g_{k}e^{-i\omega_{k}t}z_{k}$ satisfies $\mathcal{M}\{\textbf{z}_{t}\}=\mathcal{M}\{\textbf{z}_{t}^{*}\}=0$ and $\mathcal{M}\{\textbf{z}_{t}\textbf{z}_{\tau}^{*}\}=C(t-\tau)\equiv\sum_{k}|g_{k}|^{2}e^{-i\omega_{k}(t-\tau)}$, where $\mathcal{M}\{...\}$ denotes the statistical mean over all the possible stochastic processes and $C(t)$ is the environmental correction function. By far, no approximation is invoked, which means the non-Markovian quantum state diffusion equation given by Eq.~(\ref{eq:eqa2}) is exact. However, Eq.~(\ref{eq:eqa2}) is difficult to solve due to the time-non-local functional derivative term~\cite{PhysRevA.98.012110,PhysRevA.98.032116}.

Fortunately, for the Lorentzian spectral density considered in this paper, $C(t)$ is an Ornstein-Uhlenbeck-type correlation function, namely $C(t)$ can be expressed as an exponential
function: $C(t)=\alpha e^{-\beta t}$ with $\alpha=\frac{1}{2}\gamma\lambda$ and $\beta=\lambda+i\Delta$. Using this particularity, the HEOM algorithm can be realized~\cite{PhysRevA.96.032125,PhysRevA.98.012110,PhysRevA.98.032116,Suess2015}. By introducing auxiliary operators $\varrho_{t}^{(m,n)}\equiv\mathcal{M}\{|\psi_{t}^{(m)}(\textbf{z}^{*})\rangle\langle\psi_{t}^{(n)}(\textbf{z}^{*})|\}$ with
\begin{equation}\label{eq:eqa3}
|\psi_{t}^{(m)}(\textbf{z}^{*})\rangle\equiv\bigg{[}\int_{0}^{t}d\tau C(t-\tau)\frac{\delta}{\delta \textbf{z}_{\tau}^{*}}\bigg{]}^{m}|\psi_{t}(\textbf{z}^{*})\rangle,
\end{equation}
the non-Markovian quantum state diffusion equation can be reexpressed as the following HEOM~\cite{PhysRevA.96.032125,PhysRevA.98.012110,PhysRevA.98.032116,Suess2015}
\begin{equation}\label{eq:eqa4}
\begin{split}
\frac{d}{dt}\varrho_{t}^{(m,n)}=&(-i\hat{H}_{\text{s}}^{\times}-m\beta-n\beta^{*})\varrho_{t}^{(m,n)}\\
&+m\alpha\hat{L}\varrho_{t}^{(m-1,n)}+n\alpha^{*}\varrho_{t}^{(m,n-1)}\hat{L}^{\dag}\\
&-\hat{L}^{\dagger\times}\varrho_{t}^{(m+1,n)}+\hat{L}^{\times}\varrho_{t}^{(m,n+1)},
\end{split}
\end{equation}
where $\hat{X}^{\times}\hat{Y}\equiv\hat{X}\hat{Y}-\hat{Y}\hat{X}$ and $\varrho_{t}^{(0,0)}$ is the reduced density operator of the probe. Similarly, the HEOM of the spin-fermion model is given by~\cite{PhysRevA.96.032125,Suess2015}
\begin{equation}\label{eq:eqa5}
\begin{split}
\frac{d}{dt}\varrho^{(m,n)}_{t}&=(-i\hat{H}_{s}^{\times}-m\beta-n\beta^{*})\varrho^{(m,n)}_{t}\\
&+\Theta(m)\alpha \hat{L}\varrho^{(m-1,n)}_{t}+\Theta(n)\alpha^{*} \varrho^{(m,n-1)}_{t}\hat{L}^{\dag}\\
&+\Big{[}(-1)^{n}\varrho^{(m+1,n)}_{t}\hat{L}^{\dagger}-\hat{L}^{\dagger}\varrho^{(m+1,n)}_{t}\Big{]}\\
&+\Big{[}(-1)^{m}\hat{L}\varrho^{(m,n+1)}_{t}-\varrho^{(m,n+1)}_{t}\hat{L}\Big{]},
\end{split}
\end{equation}
where $\Theta(x)\equiv x~\text{mod}~2$ and the auxiliary operators are defined as $\varrho_{t}^{(m,n)}=\mathcal{M}\{|\psi_{t}^{(m)}(\textbf{z}^{*})\rangle\langle\psi_{t}^{(n)}(-\textbf{z}^{*})|\}$.

The initial-state conditions of the auxiliary operators are $\varrho^{(0,0)}_{t}=\varrho_{\text{s}}(0)$ and $\varrho^{(m>0,n>0)}_{t}=0$. In numerical simulations, we need to truncate the hierarchical equations. In practice, we choose a sufficiently large integer $N$ and set $\varrho^{(m,n)}_{t}$ with $m+n>N$ to be zero. Then, $\varrho^{(m,n)}_{t}$ with $m+n\leq N$ form a closed set of ordinary differential equations which can be numerically solved by using the well-developed Runge-Kutta algorithm.

\section*{Appendix B: Zwanzig-Nakajima master equation}

In the special case $\chi=1$, $\hat{H}$ reduces to the standard spin-boson model, whose reduced dynamics can be captured by the famous Zwanzing-Nakajima master equation~\cite{10.1143/PTP.20.948,doi:10.1063/1.1731409}
\setcounter{equation}{0}
\renewcommand\theequation{B\arabic{equation}}
\begin{equation}\label{eq:eqb1}
\frac{\partial}{\partial t}\varrho_{\mathrm{s}}(t)=-i\mathcal{\hat{L}}_{\mathrm{s}}\varrho_{\mathrm{s}}(t)-\int_{0}^{t}d\tau\hat{\Sigma}(t-\tau)\varrho_{\mathrm{s}}(\tau),
\end{equation}
where we have used the Born approximation, $\mathcal{\hat{L}}_{\mathrm{x}}\mathcal{\hat{O}}=\hat{H}_{\mathrm{x}}^{\times}\mathcal{\hat{O}}$ with $\mathrm{x}=\mathrm{s},\mathrm{e},\mathrm{i}$ are the Liouvillian operators corresponding to the probe, the environment and the probe-environment interaction Hamiltonian, respectively. Here, $\hat{\Sigma}(t)$ is the self-energy super-operator. If the probe-environment coupling is weak, $\hat{\Sigma}(t)$ can be approximately written as~\cite{PhysRevB.71.035318,PhysRevB.79.125317}
\begin{equation}\label{eq:eqb2}
\hat{\Sigma}(t)\simeq\mathrm{Tr}_{\mathrm{e}}\Big{[}\mathcal{\hat{L}}_{\mathrm{i}}e^{-it(\mathcal{\hat{L}}_{\mathrm{s}}+\mathcal{\hat{L}}_{\mathrm{e}})}\mathcal{\hat{L}}_{\mathrm{i}}\varrho_{\mathrm{e}}\Big{]}.
\end{equation}

\begin{figure}
\centering
\includegraphics[angle=0,width=0.475\textwidth]{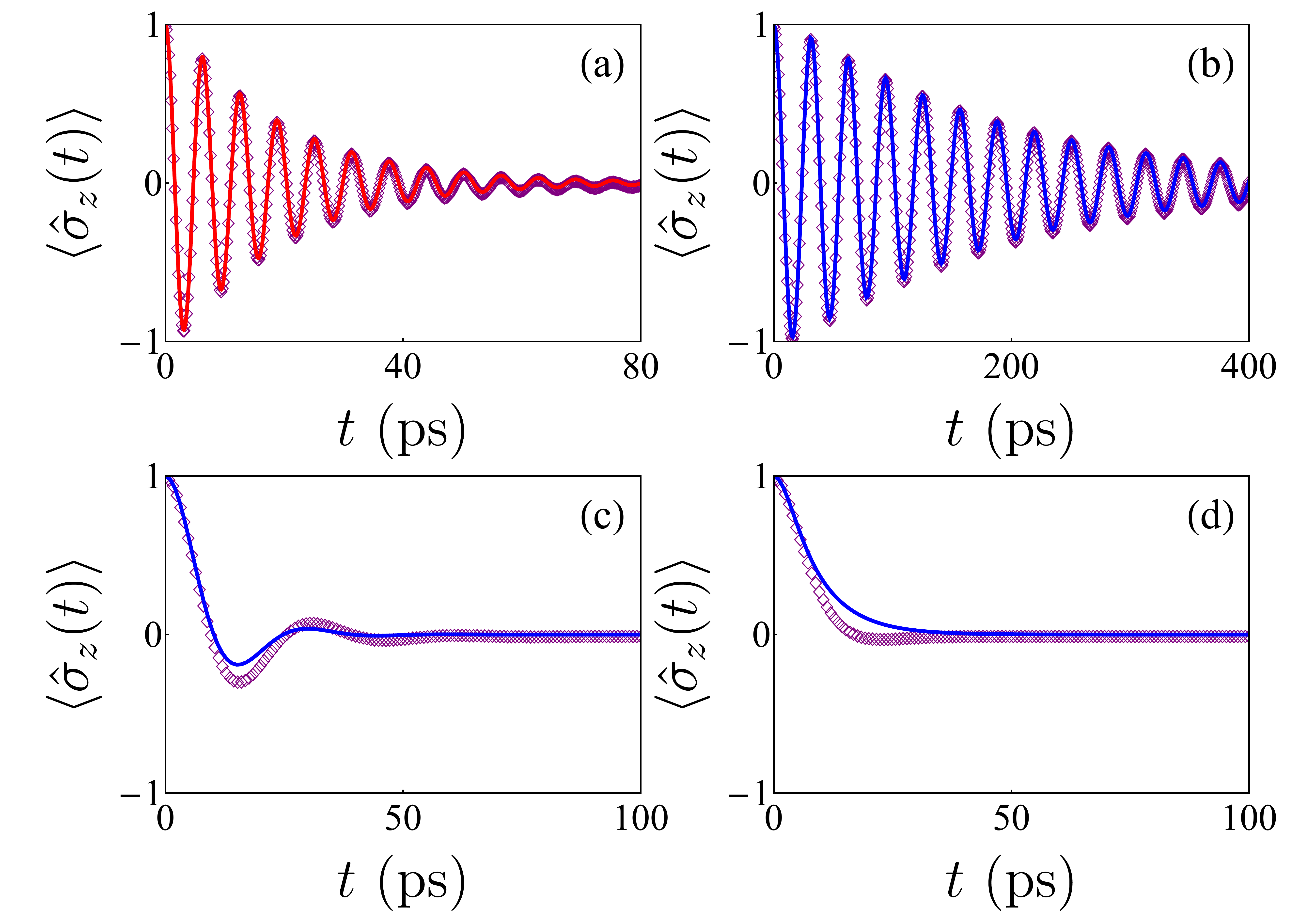}
\caption{The dynamics of the population difference with (a) $\chi=0$, $\gamma=0.1~\mathrm{cm}^{-1}$ and $\Delta=1~\mathrm{THz}$, (b) $\chi=1$, $\gamma=0.01~\text{cm}^{-1}$ and $\Delta=0.2~\text{THz}$, (c) $\chi=1$, $\gamma=0.1~\text{cm}^{-1}$ and $\Delta=0.2~\text{THz}$, (d) $\chi=1$, $\gamma=0.2~\text{cm}^{-1}$ and $\Delta=0.2~\text{THz}$. The red solid line represents the exactly analytical from the RWA treatment, the blue solid lines are obtained by the Zwanzig-Nakajima master equation approach, and the purple diamonds are purely numerical results predicted by the HEOM method. The other parameters are $\lambda=5\gamma$ and $\phi=\pi/4$.}\label{fig:fig4}
\end{figure}

In the Bloch representation, the Zwanzig-Nakajima master equation can be expressed as $\partial_{t}\vec{r}(t)=\mathfrak{S}(t)\diamond\vec{r}(t)$, where $\diamond$ denotes the convolution and
\begin{equation}\label{eq:eqb3}
\mathfrak{S}(t)=\left[
                       \begin{array}{ccc}
                         -\mathcal{A}(t) & 0 & 0 \\
                         0 & -\mathcal{B}(t) & -\Delta\delta(t) \\
                         0 & \Delta\delta(t) & 0 \\
                       \end{array}
                     \right],
\end{equation}
with $\mathcal{A}(t)=4\cos(\Delta t)C(t)$, $\mathcal{B}(t)=4C(t)$. By means of the Laplace transform, one can find
\begin{equation}\label{eq:eqb4}
\begin{split}
\tilde{r}_{i}(\zeta)\equiv&\int_{0}^{\infty}dtr_{i}(t) e^{-\zeta t}=\sum_{j}\mathfrak{F}_{ij}(\zeta)r_{j}(0),
\end{split}
\end{equation}
where $i,j=x,y,z$. The non-vanishing terms of $\mathfrak{F}_{ij}(\lambda)$ are given by
\begin{equation}\label{eq:eqb5}
\mathfrak{F}_{xx}(\zeta)=\Big{[}\zeta+\mathcal{\tilde{A}}(\zeta)\Big{]}^{-1},
\end{equation}
\begin{equation}\label{eq:eqb6}
\mathfrak{F}_{yy}(\zeta)=\bigg{[}\zeta+\mathcal{\tilde{B}}(\zeta)+\frac{\Delta^{2}}{\zeta}\bigg{]}^{-1},
\end{equation}
\begin{equation}\label{eq:eqb7}
\mathfrak{F}_{zz}(\zeta)=\Big{[}1+\zeta^{-1}\mathcal{\tilde{B}}(\zeta)\Big{]}\mathfrak{F}_{yy}(\zeta),
\end{equation}
\begin{equation}\label{eq:eqb8}
\mathfrak{F}_{yz}(\zeta)=-\mathfrak{F}_{zy}(\zeta)=-\Delta\zeta^{-1}\mathfrak{F}_{yy}(\zeta).
\end{equation}
Then, for an arbitrary given initial state, the reduced dynamics of Bloch vector $\vec{r}(t)$ can be completely determined by Eq.~(\ref{eq:eqb4}) with the help of inverse Laplace transform.

\section*{Appendix C: the RWA case}

In the special case $\chi=0$, the RWA completely removes the counter-rotating-wave terms in $\hat{H}$, resulting in the total excitation number operator $\hat{\mathcal{N}}=\hat{\sigma}_{+}\hat{\sigma}_{-}+\sum_{k}\hat{c}_{k}^{\dagger}\hat{c}_{k}$ being a constant of motion. This character leads to the spin-boson model and the spin-fermion model share the same reduced dynamical
behavior in the RWA case~\cite{PhysRevA.96.032125}. As displayed in Refs.~\cite{PhysRevA.96.032125,Breuer}, in the basis of $\{|+\rangle,|-\rangle\}$, the exactly analytical expression of $\varrho_{\mathrm{s}}(t)$ can be expressed as
\setcounter{equation}{0}
\renewcommand\theequation{C\arabic{equation}}
\begin{equation}\label{eq:eqc1}
\varrho_{\mathrm{s}}(t)=\left[
          \begin{array}{cc}
            \varrho_{++}(0)\mathcal{G}_{t}^{2} & \varrho_{+-}(0)\mathcal{G}_{t}e^{-i\Delta t} \\
            \varrho_{-+}(0)\mathcal{G}_{t}e^{i\Delta t} & 1-\varrho_{++}(0)\mathcal{G}_{t}^{2} \\
          \end{array}
        \right],
\end{equation}
where
\begin{equation}\label{eq:eqc2}
\mathcal{G}_{t}=\exp\Big{(}-\frac{1}{2}\lambda t\Big{)}\bigg{[}\cosh\Big{(}\frac{1}{2}\Omega t\Big{)}+\frac{\lambda}{\Omega}\sinh\Big{(}\frac{1}{2}\Omega t\Big{)}\bigg{]},
\end{equation}
is the decoherence factor and $\Omega\equiv\sqrt{\gamma^{2}-2\gamma\lambda}$.

With Eq.~(\ref{eq:eqc1}) at hand, the analytical expressions of CFI and QFI can be easily computed. Assuming the initial state parameter $\phi$ is chosen as $\phi=\pi/4$, we find
\begin{equation}\label{eq:eqc3}
\mathcal{F}_{\text{C}}^{\text{RWA}}=\frac{t^{2}\mathcal{G}_{t}^{2}\sin^{2}(\Delta t)}{1-\mathcal{G}_{t}^{2}\cos^{2}(\Delta t)},
\end{equation}
and $\mathcal{F}_{\text{Q}}^{\text{RWA}}=t^{2}\mathcal{G}_{t}^{2}$. One can easily check that $\mathcal{F}_{\text{Q}}^{\text{RWA}}\geq\mathcal{F}_{\text{C}}^{\text{RWA}}$, due to the fact that $0\leq\mathcal{G}_{t}\leq 1$. Moreover, in the ideal noiseless case, we have $\mathcal{G}_{t}=1$, which leads to $\mathcal{F}_{\text{C}}^{\text{ideal}}=\mathcal{F}_{\text{Q}}^{\text{ideal}}=t^{2}$. This result means the selected measurement scheme is the optimal one in the noise-free situation.

In Fig.~\ref{fig:fig4} (a), we display the the dynamics of the population difference $\langle \hat{\sigma}_{z}(t)\rangle\equiv\text{Tr}_{\text{s}}[\hat{\sigma}_{z}\varrho_{\text{s}}(t)]$ from the purely numerical HEOM method. In the especial case $\chi=0$, one can see our numerical results are completely consistent with that of the exactly analytical RWA result $\langle \hat{\sigma}_{z}(t)\rangle_{\text{RWA}}=\mathcal{G}_{t}\cos(\Delta t)$. For the case $\chi=1$, as long as the probe-environment coupling is not too strong, our HEOM results are in good agreement with the predictions from the Zwanzig-Nakajima master equation approach (see Fig.~\ref{fig:fig4}(b)-(d)). However, the HEOM results are believed to be more reliable, because the Zwanzig-Nakajima master equation approach neglects the higher-order probe-environment coupling terms, which is invalid in non-Markovian and strong-coupling regimes. These results verify the feasibility and the validity of the HEOM method.

\bibliography{reference}

\end{document}